\documentclass[10pt,conference,a4paper]{IEEEtran}
\usepackage{times,amsmath,epsfig}
\title{Spectrum Quietness Metrics for Radio Astronomy}
\author{%
{Aaron Chippendale{\small $~^{\#1}$}, Kjetil Wormnes{\small $~^{*2}$}}%, Third Author{\small $~^{\#3}$} }%
\vspace{1.6mm}\\
\fontsize{10}{10}\selectfont\itshape
$~^{\#}$CSIRO Astronomy and Space Science\\
PO Box 76, Epping, NSW 1710, Australia\\
\fontsize{9}{9}\selectfont\ttfamily\upshape
$~^{1}$Aaron.Chippendale@csiro.au%\\
\vspace{1.2mm}\\
\fontsize{10}{10}\selectfont\rmfamily\itshape
$~^{*}$European Space Research and Technology Centre\\
Keperlaan 1, 2200 AG Noordwijk, Netherlands\\
\fontsize{9}{9}\selectfont\ttfamily\upshape
$~^{2}$Kjetil.Wormnes@esa.int
}
\begin{document}
\maketitle
\begin{abstract} 
We review metrics to assess the radio quietness of sites used for radio astronomy.  Concise metrics are needed to compare candidate sites for new telescopes, to monitor the quality of existing sites, and to design telescopes to work well at a given site.  Key points of assessment are the receiver dynamic range required for the strongest interferers and the expected fraction of spectrum available for sensitive astronomical measurements.  We propose three metrics: (1) total radio frequency interference power, (2) interference-to-noise power ratio and (3) time-frequency occupancy.  Box plots of these metrics summarise large quantities of information, highlight expected ranges of interfering signal properties, and aid comparisons of sites and other factors of interest.  We provide examples for Square Kilometre Array phase one deployment in Australia based on measurements made for the selection of this site.  The Square Kilometre Array will be the largest radio telescope in the world.
%We review metrics and survey strategies for assessing the radio quietness of sites used for radio astronomy.  These metrics may be used to differentiate candidate sites for new instruments, to monitor the quality of a given site over time, and to develop performance requirements for instrument design.  Key points of assessment are the receiver dynamic range required to accommodate the strongest signals, the expected fraction of spectrum available for sensitive astronomical measurements, and the nature of fluctuations and/or trends in spectrum usage and propagation conditions.  This paper discusses applications of these metrics and survey strategies to generating the necessary information for radio quiet site selection, receiver design, observation planning, and the monitoring and preservation of radio quietness for sensitive radio astronomy measurements.
\end{abstract}

% NOTE keywords are not used for conference papers so do not populate them
%\begin{keywords}
%ignore
%\end{keywords}
%
\section{Introduction}
Our goal is to assess the impact of man-made radio frequency interference (RFI) on observatory viability and efficiency via as few numbers and plots as possible.  Carefully presented summary metrics can aid efficient and fair comparison of candidate telescope sites and identify important trends at existing sites.  RFI surveys produce very large data sets that prohibit inspection of every spectrum.  Studying individual spectra may be necessary to understand fine details.  However, a high-level metric based analysis should be made first to identify where such detailed effort is best spent.

Concise metrics are required to summarise the effects of RFI on (1) the viability of telescope receiver hardware and (2) availability of spectrum for sensitive astronomy.  Availability of spectrum for sensitive astronomical measurements is the key issue for scientific efficiency of an observatory.  However, this should not be considered independently of the viability and cost of RFI-robust receivers and signal processing.  

We propose a high-level analysis based on summary statistics of total RFI power, interference-to-noise power ratio (INR), and time-frequency occupancy.  Statistics of the first two metrics summarise the range of RFI powers that receivers must cope with at a given site.  Good results for these are required so that a practical and economical telescope can be built to realise the otherwise theoretical spectrum availability indicated by the third metric.

\begin{figure}[b!]
	\centerline{\psfig{figure=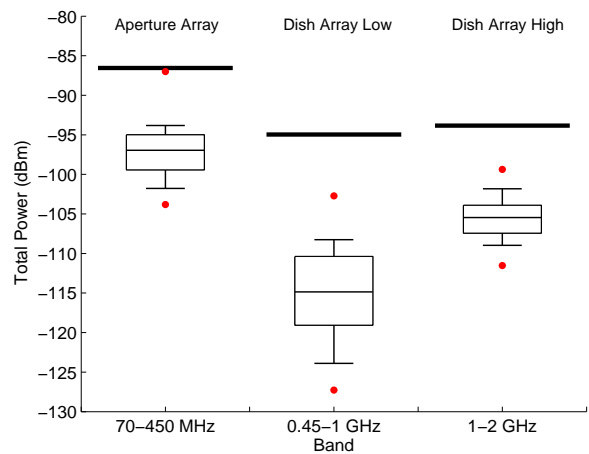,height=64.54mm} }
	\caption{Box plot summarising total RFI power detected in SKA Phase I bands at the Australian SKA site.  Boxes extend from the $25^{\text{th}}$ to $75^{\text{th}}$ percentiles and are broken by a horizontal line at the median.  Whiskers extend to 10$^{\text{th}}$ and 90$^{\text{th}}$ percentiles, and dots mark maxima and minima.  These statistics are calculated over estimates of total RFI power from each of the 853 usable 60~s integration spectra.  Thickened lines show expected receiver system noise power $P_{sys} = k\overline{T}_{sys}B$ in each band.   }
	\label{fig:boxplot}
\end{figure}

\emph{Total RFI Power} statistics help engineers design analog receiver front ends with sufficient headroom.  These front ends must operate linearly for the majority of expected RFI power levels.  Box plots of total RFI power like Fig. \ref{fig:boxplot} neatly summarise the typical range of total RFI power and additionally highlight the strongest powers that may damage a receiver.  Engineers can quickly gauge whether a design has sufficient headroom by adding their own power limit lines for receiver damage and input 1~dB compression to such plots.

\emph{Interference-to-Noise Power Ratio} (INR) statistics help engineers design digital back ends with sufficient headroom.  INR is the ratio of total power in detected RFI signals to total noise power expected from an astronomy receiver in a specified band.  Typical range and and extreme values of INR can be read from Fig. \ref{fig:boxplot} by comparing the box plots to the thickened system noise power lines.  This helps engineers assess the number of bits required to keep digital errors below specified thresholds.  

\begin{figure}[t!]
	\centerline{\psfig{figure=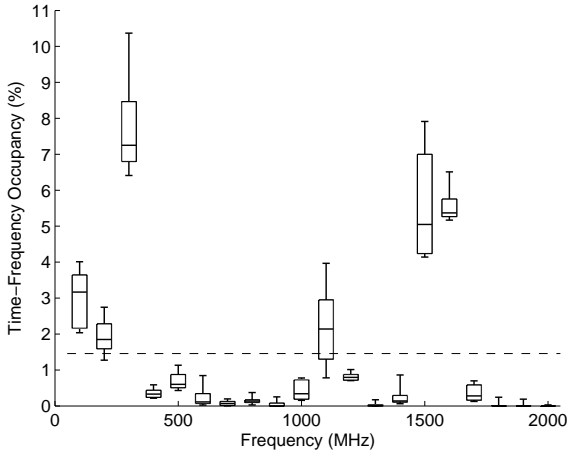,height=64.54mm} }
	\caption{Time-frequency occupancy in 100~MHz bands at the Australian SKA site.  This is the percentage of 27.4~kHz$\ \times \ $60~s data points affected by RFI in each 100~MHz$\ \times \ $2~hr high-sensitivity spectrogram.  Variations over all data from 8 spectrograms covering 4 directions $\ \times \ $ 2 polarisations are shown by boxes extending from the 25$^{\text{th}}$ to 75$^{\text{th}}$ percentiles that are split by horizontal lines at median values.  Whiskers extend to 10$^{\text{th}}$ and 90$^{\text{th}}$ percentiles.  The dashed line shows mean time-frequency occupancy for all data.}
	\label{fig:occupancy}
\end{figure}

\emph{Time-Frequency Occupancy} statistics help astronomers estimate the overall efficiency of an observatory.  They indicate what fraction of data is affected by RFI at a particular sensitivity threshold.  This fraction of data cannot be used for science without invoking additional signal processing for RFI mitigation.  Fig. \ref{fig:occupancy} shows an example.

\section{Measuring RFI Power}
\subsection{Equivalent Noise Power}
We measure power received by a radio telescope in terms of an equivalent noise temperature $T$.  This means the power measured with the antenna is equal to the power measured when the antenna is replaced by a matched resistor at physical temperature $T$.  Power received by the telescope is represented by the thermal noise power of this hypothetical matched resistor $P = kT\Delta f$
where $\Delta f$ is the measurement bandwidth and $k$ is Boltzmann's constant. We assume that the telescope is rarely pointed directly at RFI sources and that RFI is typically received through 0~dBi telescope sidelobes.   %Power measured by a low-gain RFI survey antenna is comparable to the power received through the sidelobes of a high-gain telescope antenna on the same site.

%If the telescope's antenna pattern is subtended by a distant object emitting thermal radiation then $T$ also corresponds closely to the physical temperature of that body.  

%We assume the telescope does not point directly at the RFI source.  Astronomy cannot be done in that configuration as typical astronomical sources are many orders of magnitude weaker than even weak RFI.  We assume RFI enters 0~dBi telescope sidelobes and therefore does not experience the large directive gain of the telescope antenna. 

\subsection{Sensitivity}
Astronomers discuss the sensitivity of telescopes with reference to the smallest noise temperature $\Delta T$ they can detect
\begin{equation}
  \Delta T = \frac{T_{sys}}{\sqrt{\Delta f\tau}}
  \label{eq:sens}
\end{equation}
with integration time $\tau$, bandwidth $\Delta f$, and significance of one standard deviation.  $\Delta T$ is proportional to the standard error in the estimated standard deviation (r.m.s. power) of the measured voltage.  For a Gaussian distributed noise voltage, $\Delta T$ reduces in inverse proportion to the square root of the number of independent measurements $\Delta f\tau$ as shown by \eqref{eq:sens}.

System equivalent noise temperature ${T_{sys}=T_{rx} + T_{sky}}$ has contributions from the receiver $T_{rx}$ and the radio-sky background $T_{sky}$.  Below 400~MHz $T_{sky}$ is dominated by the Galactic radiation and can be modelled by ${T_{sky} = 60\lambda^{2.55}\ \text{K}}$ \cite{Bregman2000}.  From approximately 1~GHz to 10~GHz a flat value of 8~K is more appropriate, arising from the cosmic microwave background of 3~K and atmospheric emission of approximately 5~K.  This suggests a simple sky noise model for receiver design purposes below 10~GHz
\begin{equation}
    T_{sky} = 
     \max \left(2.78f_{\text{GHz}}^{-2.55} , 8\right) \ \text{K}.
\end{equation}

\subsection{Raw Data}
Most RFI data come in the form of dynamic spectra (spectrograms) or raw voltages sampled at the Nyquist rate.    Spectrograms are good for characterising the quasi-stationary RFI background of TV/radio broadcasts and radio telecommunications.  With appropriate calibration, spectrograms can be time-averaged in software to yield high sensitivity.  Recording raw voltages is efficient for detecting and analysing fast-transient RFI such as radar chirps and aviation transponder pulses.  Efficiency is increased further by recording voltage samples to disk only when a power threshold is exceeded.

%Spectrograms or Dynamic Spectra $P(f_i, t_j)$ are matrices of power measurements with dimension $N$ spectral channels by $M$ measurement epochs.  They have a characteristic spectral resolution $\Delta f$, measurement cadence $\Delta t$, and integration time $\tau$.  With appropriate calibration, many such power spectra can be averaged in software to achieve high sensitivity.  %Traditionally, spectrograms were measured with swept receivers that inefficiently measured one frequency channel at a time for a low integration time ($\tau<1 \ \text{ms}$).  This resulted in low duty-cycle $\delta=\tau/\Delta t$ and low sensitivity.  More recently, fast Fourier transform and poly-phase filterbank integrating spectrometers have been used to measure power simultaneously across large instantaneous bandwidths with longer integration times ($\tau \geq 1 \ \text{s}$) and 100\% duty-cycle ($\delta=1$).  

Contemporary spectrum surveys of radio astronomy sites might include:
\begin{enumerate}
  \item  \emph{low-sensitivity spectrograms} with low integration ($\tau<1$~sec), peak detection, and moderate instantaneous bandwidth  scanned rapidly over multi-GHz bands for long periods to measure variations in dominant RFI signals as spectrum usage and propagation conditions vary over seconds to years;
  \item  \emph{high-sensitivity spectrograms} with long total integration ($\tau>1$~hr) and large instantaneous bandwidths to detect weak RFI signals that may limit spectrum use for sensitive astronomical measurements; and
  \item \emph{raw-voltage recordings} at the Nyquist rate ($<0.5$~ns) for broad instantaneous bandwidths to characterise impulsive and/or sporadic interference to which (1) and (2) are insensitive.
\end{enumerate}

Data used for examples in this paper come from the selection process for the world's largest radio telescope, the Square Kilometre Array \cite{Dewdney2009}.  High-sensitivity spectrograms for both the Australian and South African SKA sites are freely available at \cite{ska2013} and their measurement and calibration are described in \cite{Boonstra2011}.  We calculate the proposed RFI metrics from all high-sensitivity spectrograms collected at the Australian SKA site in 2010/2011.  Each measurement covered 50~MHz to 2,050~MHz with 27.4 kHz spectral resolution, 60~s integration periods, and 2~hr total integration.  Eight measurements were made to cover all combinations of four cardinal antenna pointings and two antenna polarisation orientations -- horizontal and vertical.

 % We don't consider the transient-mode SKA data.  They were recorded via 1~$\mathrm{\mu}$S integration spectra streamed continuously for half-second bursts between 10~minute download intervals.  These sparse untriggered recordings are not useful for assessing the population of transient signals. 

%Raw spectrograms were calibrated into Watts using a calibrated noise diode as an absolute power standard, converted to an estimate of flux density at the RFI survey antenna location in ${\text{W}.\text{m}^{-2}}$ assuming all signals are received at the peak of the RFI survey antenna's pattern.  Fluxes are then converted to power at the terminals of a hypothetical radio telescope antenna assuming reception through 0~dBi sidelobes.  A comparison load switch was used to allow long-term integration to achieve high sensitivity in the face of fluctuating receiver gains and additive spectral artifacts generated by the back end.

Detected RFI power is calculated by subtracting an estimate of the noise power in RFI free channels from each spectrum and then thresholding the result.  The noise-power baseline was estimated by median filtering the spectra.  We used a threshold of 6$k\Delta T\Delta f$ representing six times the smallest detectable signal.  This method is insensitive to RFI broader than 274~kHz, which is half the width of the median filter.

The authors of the present paper were part of a CSIRO team that delivered the comparison load switch, the analog band-limiting filterbank, the ADC, the poly-phase filterbank spectrometer, and prototype calibration and analysis algorithms for the SKA site selection survey work. 

\section{Assessing RFI Quietness with Metrics}

\subsection{Total RFI Power and Analog Headroom}
\label{sec:total-rfi-power}
The first priority is to summarise the strongest RFI powers experienced on a site.  This helps determine if it is practical and affordable to make receivers with sufficient headroom for the site.  Astronomy receivers must handle the maximum total in-band RFI power $P_{totRFI}$ without being damaged.   They should also handle the $90^{\text{th}}$ percentile $P_{totRFI}$ without introducing broad-band errors via gain compression and intermodulation.  Otherwise there would be significant impact to scientific observations for 10\% of the time. 

Levels of dominant RFI are neatly summarised by box plots of total RFI power in bands of interest.  Key bands of interest are the antenna element bandwidth, analog front-end bandwidth, and the band-limited sampled bandwidth.  Total RFI power in a given band can be calculated directly from raw voltage samples, or estimated by summing power in RFI-occupied channels over a spectrogram.  It is possible to correct for the fact that multiple modulated narrow-band RFI sources are unlikely to be received in-phase \cite{Perley2004}.  We, however, stick to the linear sum as a conservative worst-case limit.

Fig. \ref{fig:boxplot} shows total RFI power box plots for preliminary SKA Phase I receiver bands defined in \cite{Dewdney2010} and summarised in Tab. \ref{tab:ska1}.  This plot clearly shows extreme values of $P_{totRFI}$ along with typical ranges.  The total RFI power was calculated by summing the power in detected RFI signals over the nominated band for each 60~s spectrum.  The reported statistics are calculated over the 853 usable 60~s spectra collected.  Fig. \ref{fig:boxplot} compares distributions of  $P_{totRFI}$ between different receiver bands, but could equally compare sites, time of day, or any other factor of interest.  Thickened horizontal lines represent expected system noise powers $P_{sys}=k\overline{T}_{sys}B$ for SKA Phase I receivers.  Band average system temperature $\overline{T}_{sys}$ is calculated by averaging $T_{sys}$ over the band of interest.

\begin{table}[b!]
\centering

    \caption{Preliminary SKA Phase I Bands \cite{Dewdney2010}}     % NOTE!  caption goes _before_ the table contents !!
    \label{tab:ska1}

    \begin{small}
    \begin{tabular}{|l|l|l|}
    \hline
    %{\bfseries } & \multicolumn{3} {c|} {\bfseries Appearance (in Times New Roman or Times} \\
    %\cline{2-4}
     {\bfseries  Band Name}         & {\bfseries Frequency}     & {\bfseries Receiver }           \\
     {\bfseries  }         & {\bfseries Range}     & {\bfseries Temp. $T_{rx}$ }           \\
     {\bfseries  }         & {\bfseries (MHz)}     & {\bfseries (K)}      \\   
    \hline
     Sparse	Aperture Array&	 70-450	& 150	\\
        \hline
     Dish	Array Low& 450-1000 &	32\\
    \hline
    Dish Array High	&	1000-2000	& 22      \\
    \hline
    \end{tabular}
    \end{small} 
\end{table}

The example data don't capture the strongest power peaks.  They are high-sensitivity spectrograms measured over just three days per antenna orientation by a spectrometer with limited headroom.  The RFI survey receiver, optimised to detect weak signals, was saturated by strong RFI on a number of occasions in some bands at both South African and Australian SKA sites.  We discarded saturated spectra from the current analysis.  This is acceptable for assessing the population of weak RFI sources in the total integration, but totally inadequate for assessing the distribution of strong RFI sources that limit receiver design.  The examples presented here are illustrations of the metrics, not complete representations of the site.  It is likely that the typical ranges of RFI power are reasonable but that the maximum levels are significantly underestimated due to limitations of the data set.  

Longer data sets collected with higher head-room receivers and peak detection are required to properly characterise peak levels of RFI on both SKA sites.  It is critical that the receiver for this component of an RFI survey has sufficient headroom to measure the strongest RFI without saturating.  This may require the use of a pre-receiver attenuator, which is acceptable for low-sensitivity measurements of strong RFI. 
% If it is necessary to characterise strong peaks and weaker signals in a single measurement setup, sensitivity lost from the attenuation can be recovered by subtracting reference measurements of a matched load from spectra and integrating for longer \cite{Rogers2008}. 

The spectra used to calculate the total interference power should be measured instantaneously across the full observation band.  The SKA site measurement spectrometer required the 50~MHz to 2,050~MHz band to be measured sequentially in 9 sub-bands. Estimating instantaneous RFI power depends on the unknown likelihood of RFI measured in different bands at different times occurring at the same time. It would be better to measure the full band of interest simultaneously with a shorter integration to assess the instantaneous distribution of total RFI power that is most relevant for system design.

\subsection{Interference-to-Noise Power Ratio and Digital Headroom}
\label{sec:inr}
Interference-to-noise power ratio $a = P_{totRFI} / P_{sys}$ is the key metric required to estimate gain error and distortion terms due to signal clipping and quantisation in digital back ends.  It is the ratio of total measured RFI power to total system noise power from an astronomy receiver.  This ratio can be read off Fig. \ref{fig:boxplot} by measuring the distance in dB of the various total RFI power statistics to the thickened system noise power line for the relevant band.  This may be more clearly presented by direct boxplots of $a$ on a dB scale.    

As $a$ exceeds unity the back-end hardware rapidly increases in cost, complexity, and effort to design. Digital engineers must carefully choose sufficient bit-resolution to accommodate the RFI power.  They must carefully balance precision and headroom through all digital signal processing.  Digital design is much simpler if $a \ll 1$ and the RFI power is buried in system noise.  If $a < 0.03$ (-15~dB) a high spectral dynamic range (approx. 60 dB) can be achieved even with 1-bit sampling and appropriate non-linear corrections \cite{Chippendale2005}. 

Quantisation and subsequent digital processing introduces significant errors when RFI exceeds design headroom.  These errors can be partitioned into a digital gain error and an additive distortion term \cite{Bendat1990}.  The digital gain error and the total power in the distortion term may be calculated via Busgang's theorem \cite{Bussgang1975,Bendat1990} following the procedure in Chapter 6 of \cite{Chippendale2009}.  Both of these error terms depend directly  on $a$.  Knowing $a$ and having a specification for maximum allowable spectral errors allows calculation of the necessary number of sampling levels.  It is then up to careful digital design to maintain the dynamic range through subsequent digital signal processing.

Estimating $a<1$ from spectrograms is biased.  Summing detected RFI to give $P_{totRFI}$ misses weak RFI just below the system noise floor $P_{sys}$.  However, a power law distribution governing the number of interferers at each power level might be determined from RFI measurements and propagation modelling. Knowing the form of this distribution and being able to extrapolate it to unmeasurably low power levels may improve estimates of $P_{totRFI}$.  This would reduce bias in estimating small values of $a$ and enable error estimates for $P_{totRFI}$ and $a$.  Power in very weak interferers, buried in the noise, may be estimated via kurtosis methods \cite{DeRoo2007}, but this involves assumptions about modulation of the RFI signals.

%Fig. \ref{fig:boxplot-ratio} shows an estimate of $a$ for the Australian SKA site.  Total RFI power was computed by summing the mean power in all channels where RFI was detected in a given trace in a given observation band. Total noise power was calculated by integrating ${T_{sys} = T_{rx} + T_{sky}}$ across the same observation band.  The same caveat exists as for Fig. \ref{fig:boxplot} in that the data were not collected with peak detection and that some spectra were discarded when the RFI survey equipment saturated.

%\begin{figure}[t!]
%	\centerline{\psfig{figure=inr-boxplot.eps,height=64.54mm} }
%	\caption{Boxplot summarising total RFI power to noise power ratio in SKA Phase I bands at the Australian SKA site.  Boxes show the range from the 25$^{\text{th}}$ to 75$^{\text{th}}$ percentiles split by a horizontal line representing median.  Whiskers extend to 10$^{\text{th}}$ and 90$^{\text{th}}$ percentiles, and dots mark minimum and maximum values.  The thick line at 0 dB represents the critical break point when the ratio $a=1$ and interference power equals receiver noise power.}
%	\label{fig:boxplot-ratio}
%\end{figure}

\subsection{Time-Frequency Occupancy and Observatory Efficiency}
\label{sec:occupancy}
Time-frequency occupancy estimates the overall fraction of RFI-affected data that would normally be discarded before science processing.  Equation \eqref{eq:sens} highlights that losing 10\% bandwidth or 10\% time has the same impact of making the observatory 10\% less efficient for continuum measurements of broadband power.  However, astronomers studying spectral lines prefer uninterrupted bandwidth and those studying transients prefer uninterrupted time.  It is difficult to define a single occupancy that represents observatory efficiency for all astronomy.  Occupancies are also sensitive to the time and frequency resolution of RFI survey data.  These should be matched to resolutions planned for telescope operation.  

Fig. \ref{fig:occupancy} shows box plots of the time-frequency occupancy in 100~MHz sub-bands at the Australian SKA site.  This is the fraction of RFI-affected 27.4~kHz$\ \times \ $60~s measurement points in each 100~MHz$\ \times \ $2~hr spectrogram.  Each box shows variations over the 8 high-sensitivity spectrograms due to direction, polarisation, and time of measurement.  Larger data sets over longer monitoring periods are required to discover what portion of variation is due to each of these factors.    

Astronomy data are considered degraded when RFI power increases measured noise temperature by 10\% of the minimum detectable noise temperature $\Delta T$ when RFI is absent \cite{Thompson1982, itu2003}.  However, $\Delta T$ for astronomy receivers is much much smaller than the sensitivity achievable with RFI survey receivers.  Further study is required to understand the relationship between occupancies at RFI survey sensitivities and actual occupancies realised by sensitive radio telescopes.  RFI survey sensitivity could also be boosted by advanced detection methods based on higher order statistics such as kurtosis.

%\begin{figure}[t!]
%	\centerline{\psfig{figure=occupancy-single-au-sa.eps,height=64.54mm} }
%	\caption{Distribution of single-trace spectral occupancy for SKA Phase I bands at Australian (AU) and South African (SA) SKA sites.  Boxes show the range from the 25$^{\text{th}}$ to 75$^{\text{th}}$ percentiles split by a horizontal line representing median.  Whiskers extend to 10$^{\text{th}}$ and 90$^{\text{th}}$ percentiles, and dots mark minimum and maximum values. }
%	\label{fig:occupancy-single}
%\end{figure}

\section{Assessing RFI Mitigation Potential}
When the metrics indicate problematic RFI, detailed analysis of dominant interferers can identify potential mitigation strategies including moving receiving bands or introducing notch filters.  Usefulness of these traditional approaches can be gauged by their impact on the metrics.

In the future, observatories can be made more efficient by active RFI mitigation techniques \cite{Kesteven2010} that cancel RFI and allow use of data that is traditionally discarded.  RFI survey strategies and metrics should be developed to assess how successful known RFI mitigation algorithms could be at a given site.  It may become important to measure the interference-to-noise ratios for individual interferers, the number of independent interferers per signal-processing channel bandwidth, and rates of RFI source motion.  This additional information  may aid calculation of the expected costs and benefits of implementing various mitigation strategies.

\section{Conclusions}
RFI metrics to guide practical receiver design are as important as metrics estimating available spectrum for astronomy assuming an ideal receiver.  These issues should not be separated.  We hope this work inspires an open discussion  of these and other metrics.  Progressing this work to a standard procedure for high-level analysis and communication of RFI data would significantly benefit the international astronomical community and the large international projects it is pursuing.  These involve many stakeholders that must share information across many disciplines to select good sites, protect good sites, and design hardware that works well at these sites.

Further measurements at both SKA sites are required to collect the information required to design successful SKA receivers.  These measurements need to be made with higher dynamic range receivers, peak detection, and triggered recording of transients.  These, along with further high-sensitivity spectrogram measurements, should be carried out over long periods to better understand variations in spectrum usage and propagation over time-scales from nanoseconds to years.

%Australia (19.S.H) and Australia (17.E.H) typically maintain a 10~dB margin most (should quantify this as a percentage) of the time for both observation bands and would typically allow the full dynamic range of a given ADC to be achieved.

% the following vfill coursely balances the columns on the last page
%\vfill
%
%\pagebreak

\section*{Acknowledgment}
Thanks to Dick Manchester, Ron Beresford and Lisa Harvey-Smith for helpful discussion.  This work uses data obtained from the Murchison Radio-astronomy Observatory (MRO), which is jointly funded by the Commonwealth Government of Australia and the State Government of Western Australia.  CSIRO manages the MRO and provides operational support to ASKAP.  We acknowledge the Wajarri Yamatji people as the traditional owners of the Observatory site.

%Kjetil Wormnes

\bibliographystyle{IEEEtran}

\bibliography{IEEEabrv,../../../references/rfi}

\begin{thebibliography}{10}
\providecommand{\url}[1]{#1}
\csname url@samestyle\endcsname
\providecommand{\newblock}{\relax}
\providecommand{\bibinfo}[2]{#2}
\providecommand{\BIBentrySTDinterwordspacing}{\spaceskip=0pt\relax}
\providecommand{\BIBentryALTinterwordstretchfactor}{4}
\providecommand{\BIBentryALTinterwordspacing}{\spaceskip=\fontdimen2\font plus
\BIBentryALTinterwordstretchfactor\fontdimen3\font minus
  \fontdimen4\font\relax}
\providecommand{\BIBforeignlanguage}[2]{{%
\expandafter\ifx\csname l@#1\endcsname\relax
\typeout{** WARNING: IEEEtran.bst: No hyphenation pattern has been}%
\typeout{** loaded for the language `#1'. Using the pattern for}%
\typeout{** the default language instead.}%
\else
\language=\csname l@#1\endcsname
\fi
#2}}
\providecommand{\BIBdecl}{\relax}
\BIBdecl

\bibitem{Bregman2000}
J.~D. {Bregman}, ``{Design Concepts for a Sky Noise Limited Low Frequency
  Array},'' in \emph{Perspectives on Radio Astronomy: Technologies for Large
  Antenna Arrays}, A.~B. {Smolders} and M.~P. {van Haarlem}, Eds., 2000, p.~23.

\bibitem{Dewdney2009}
P.~Dewdney, P.~Hall, R.~Schilizzi, and T.~Lazio, ``The {S}quare {K}ilometre
  {A}rray,'' \emph{Proceedings of the IEEE}, vol.~97, no.~8, pp. 1482 --1496,
  Aug. 2009.

\bibitem{ska2013}
\BIBentryALTinterwordspacing
Site raw data. {SKA} Organisation. Accessed: 28/02/2013. [Online]. Available:
  \url{http://www.skatelescope.org/site-raw-data/}
\BIBentrySTDinterwordspacing

\bibitem{Boonstra2011}
\BIBentryALTinterwordspacing
A.~J. Boonstra and R.~P. Millenaar, ``{SKA} site spectrum monitoring,
  measurement program and data processing,'' {ASTRON}, Tech. Rep., 2011.
  [Online]. Available:
  \url{http://www.skatelescope.org/the-location/site-documentation/}
\BIBentrySTDinterwordspacing

\bibitem{Perley2004}
R.~Perley, ``Headroom requirements for the {EVLA},'' NRAO, EVLA Memo~82, Aug.
  2004.

\bibitem{Dewdney2010}
P.~Dewdney, J.-G. bij~de Vaate, K.~Cloete, A.~Gunst, D.~Hall, R.~McCool,
  N.~Roddis, and W.~Turner, ``{SKA} phase 1: Preliminary system description,''
  SKA Memo 130, 2010.

\bibitem{Chippendale2005}
A.~P. Chippendale, ``Effects of interference on the {ATNF} cosmolgical
  reionization experiment at mileura,'' in \emph{Proceedings of the XXVIIIth
  URSI General Assembly}, New Delhi, 2005.

\bibitem{Bendat1990}
J.~S. Bendat, \emph{Nonlinear System Analysis and Identification from Random
  Data}.\hskip 1em plus 0.5em minus 0.4em\relax Wiley, 1990.

\bibitem{Bussgang1975}
J.~J. Bussgang, \emph{Nonlinear Systems}.\hskip 1em plus 0.5em minus
  0.4em\relax Dowden, Hutchinson and Ross, 1975, ch. Cross-correlation
  functions of amplitude distorted {G}aussian inputs.

\bibitem{Chippendale2009}
A.~P. {Chippendale}, ``{Detecting cosmological reionization on large scales
  through the 21 cm HI line},'' Ph.D. dissertation, University of Sydney, 2009.

\bibitem{DeRoo2007}
R.~D. De~Roo, S.~Misra, and C.~S. Ruf, ``Sensitivity of the kurtosis statistic
  as a detector of pulsed sinusoidal {RFI},'' \emph{{IEEE} Transactions on
  Geoscience and Remote Sensing}, vol.~45, pp. 1938--1946, Jul. 2007.

\bibitem{Thompson1982}
A.~Thompson, ``The response of a radio-astronomy synthesis array to interfering
  signals,'' \emph{Antennas and Propagation, IEEE Transactions on}, vol.~30,
  no.~3, pp. 450 -- 456, May 1982.

\bibitem{itu2003}
\emph{Recommendation ITU-R RA.769-2: Protection criteria used for radio
  astronomical measurements}, ITU Std. ITU-R RA.769-2, May 2003.

\bibitem{Kesteven2010}
M.~{Kesteven}, ``Overview of {RFI} mitigation methods in existing and new
  systems (invited),'' in \emph{RFI2010 RFI Mitigation Workshop}, Radionet --
  $7^{\text{th}}$ Framework Programme.\hskip 1em plus 0.5em minus 0.4em\relax
  Groningen: Proceedings of Science, Mar 2010, {PoS(RFI2010)007}.

\end{thebibliography}

\end{document}